\documentclass[prl,superscriptaddress,nofootinbib,floats,onecolumn,longbibliography]{revtex4-2}

\usepackage{amssymb,amsmath}
\DeclareMathAlphabet\mathbfcal{OMS}{cmsy}{b}{n}
\usepackage{cmap}
\usepackage{graphicx} 
\usepackage{bm}
\usepackage{color}
\usepackage{blindtext}
\usepackage{hyperref}
\usepackage{listings}
\usepackage{booktabs}
\usepackage{multirow}
\usepackage{amsmath}
\usepackage{mathrsfs}
\usepackage{braket}
\usepackage{mathtools}

\usepackage{dsfont}
\usepackage{tabularx}

\begin{document}


\title{Topological water-wave structures manipulating particles}

\author{Bo Wang$^\dagger$}

\affiliation{State Key Laboratory of Surface Physics, Key Laboratory of Micro- and Nano-Photonic Structures (Ministry of Education), and Department of Physics, Fudan University, Yangpu District, Shanghai, 200433, China}
\affiliation{Henan Key Laboratory of Quantum Materials and Quantum Energy, School of Quantum Information Future Technology, Henan University, Zhenzhou, 450046, China}
\affiliation{Institute of Quantum Materials and Physics, Henan Academy of Sciences, Zhengzhou, 450046, China}


\author{Zhiyuan Che$^\dagger$}

\affiliation{State Key Laboratory of Surface Physics, Key Laboratory of Micro- and Nano-Photonic Structures (Ministry of Education), and Department of Physics, Fudan University, Yangpu District, Shanghai, 200433, China}


\author{Cheng Cheng}

\affiliation{State Key Laboratory of Surface Physics, Key Laboratory of Micro- and Nano-Photonic Structures (Ministry of Education), and Department of Physics, Fudan University, Yangpu District, Shanghai, 200433, China}

\author{Caili Tong}

\affiliation{Henan Key Laboratory of Quantum Materials and Quantum Energy, School of Quantum Information Future Technology, Henan University, Zhenzhou, 450046, China}

\author{Lei~Shi$^*$}

\affiliation{State Key Laboratory of Surface Physics, Key Laboratory of Micro- and Nano-Photonic Structures (Ministry of Education), and Department of Physics, Fudan University, Yangpu District, Shanghai, 200433, China}

\author{Yijie Shen$^*$}

\affiliation{Centre for Disruptive Photonic Technologies, School of Physical and Mathematical Sciences \& The Photonics Institute, Nanyang Technological University, Singapore 637371, Singapore}

\affiliation{School of Electrical and Electronic Engineering, Nanyang Technological University, Singapore 639798, Singapore}

\author{Konstantin Y. Bliokh$^*$}

\affiliation{Theoretical Quantum Physics Laboratory, Cluster for Pioneering Research, RIKEN, Wako-shi, Saitama 351-0198, Japan}

\affiliation{Centre of Excellence ENSEMBLE3 Sp. z o.o., 01-919 Warsaw, Poland}

\affiliation{Donostia International Physics Center (DIPC), Donostia-San Sebasti\'{a}n 20018, Spain}

\author{Jian Zi$^*$}

\affiliation{State Key Laboratory of Surface Physics, Key Laboratory of Micro- and Nano-Photonic Structures (Ministry of Education), and Department of Physics, Fudan University, Yangpu District, Shanghai, 200433, China}


\begin{abstract}
Topological wave structures, such as vortices and skyrmions, appear in a variety of quantum and classical wave fields, including optics and acoustics. In particular, optical vortices have found numerous applications ranging from quantum information to astrophysics.
Furthermore, both optical and acoustic structured waves are crucial for manipulation of small particles, from atoms to macroscopic biological objects. Here we report on the controllable generation of topological structures -- wave vortices, skyrmions, and polarization M\"{o}bius strips -- in interfering gravity water waves. Most importantly, we demonstrate efficient manipulation of subwavelength and wavelength-order floating particles with topologically structured water waves. This includes trapping of the particles in the high-intensity field zones, as well as controllable orbital and spinning motions due to the orbital and spin angular momenta of water waves. Our results reveal the water-wave counterpart of optical and acoustic manipulations, which paves the avenue for applications in hydrodynamics and microfluidics. 
\end{abstract}

{
\let\clearpage\relax
\maketitle
}


\def\thefootnote{$\dagger$}\footnotetext{These authors contributed equally to this work}

\def\thefootnote{*}\footnotetext{Email: lshi@fudan.edu.cn, yijie.shen@ntu.edu.sg, kostiantyn.bliokh@riken.jp, jzi@fudan.edu.cn}

\vspace{-0.5cm}

\section{Introduction}

Linear plane waves, i.e., sinusoidal oscillations propagating in one direction, are characterized by a few key parameters: amplitude, phase, frequency, wavevector, and polarization (in the case of vector waves). These are equally relevant to acoustic, electromagnetic, quantum, or hydrodynamical waves. However, when several plane waves interfere, the resulting complex or {\it structured} field becomes rather complicated, so that its amplitude, phase, and polarization can vary arbitrarily from point to point \cite{Rubinsztein2016JO, Bliokh2023JO}, even being restricted by the corresponding wave equations. For such complex wavefields, {\it topological} properties, robust with respect to small perturbations, become relevant. 

Topological wave forms, such as phase singularities (wave vortices) \cite{Nye1974, Soskin2001PO, Dennis2009PO, Shen2019LSA, Guo2022JAP, Bliokh2017PR}, polarization singularities and M\"{o}bius strips \cite{Nye1987,  Dennis2009PO,  Freund2010OC, Bauer2015S, BAD2019, Muelas2022PRL}, as well as skyrmions and merons \cite{Tsesses2018S, Du2019NP, Dai2020N, Deng2022NC, Ge2021PRL, Muelas2022PRL, Cao2023SA, Shen2024NP}, play important roles in various areas of modern wave physics. In addition to the topological robustness, these structures exhibit remarkable {\it dynamical} properties: for example, wave vortices carry quantum-mechanical-like orbital angular momentum (OAM) \cite{Allen_book,Andrews_book, Guo2022JAP, Bliokh2017PR}. That is why vortex and, generally, structured waves have enormously advanced optical and acoustic manipulations of small particles \cite{Grier2003N, Gao2017LSA, Ozcelik2018NM, Dholakia2020NRP}. 

Linear water-surface waves, including gravity and capillary waves, provide the most accessible and widely known example of classical waves. Surprisingly, systematic studies of controlled structured water waves with nontrivial topological and dynamical properties, similar to optical and acoustic structures waves, have started only recently \cite{Bacot2016NP, Filatov2016PRL, Francois2017NC, Rozenman2019Fluids, Han2022PRL, Bliokh2022SA, Che2024PRL, Smirnova2024PRL, Zhu2024NRP}. In particular, water-wave vortices with different topological charges, skyrmions, and polarization M\"{o}bius strips were described theoretically in \cite{Smirnova2024PRL, Bliokh2021POF}. 

\begin{figure*}[t]
\centering
\includegraphics[width=0.8\linewidth]{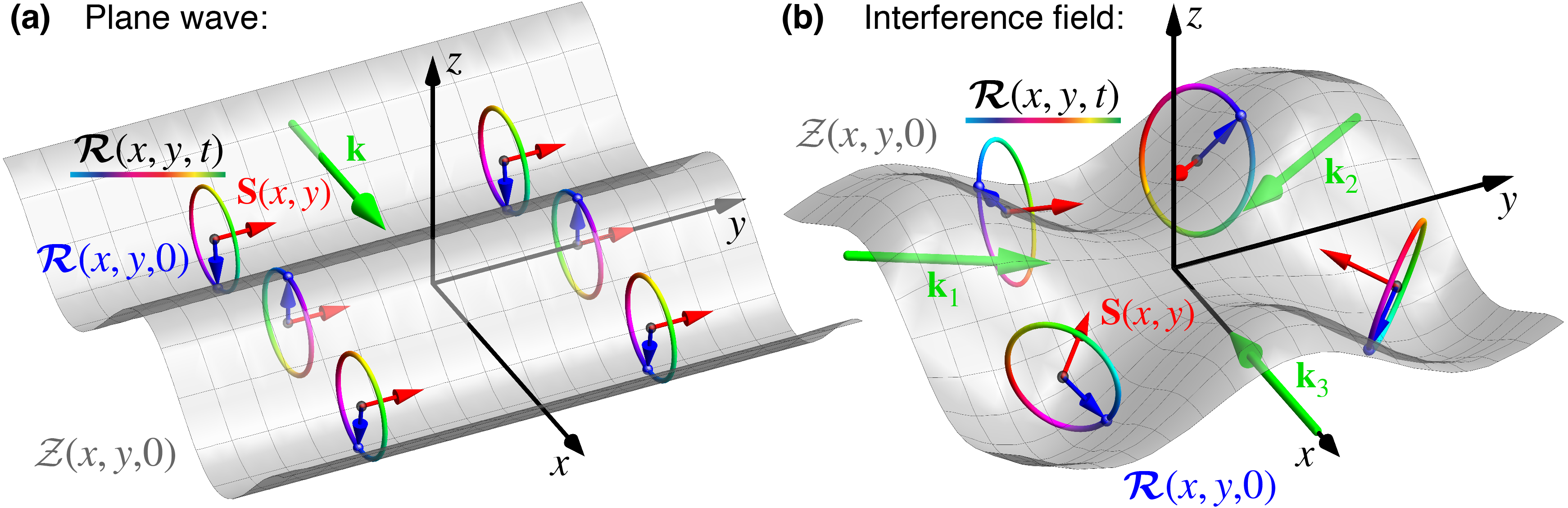}
\caption{{\bf Linear water waves and their main local characteristics.} Shown are: the wavevectors ${\bf k}$ (green), 
the wave-perturbed water surface $z= {\mathcal Z} (x,y,t=0)$ (gray), the local 3D displacement of the water-surface particles, ${\mathbfcal R}(x,y,0)$ (blue), the local elliptical trajectories (`polarizations') of the water-surface particles traced by ${\mathbfcal R}(x,y,t)$ (rainbow colors indicating the periodic evolution in $t$), and the corresponding time-averaged spin angular momentum density ${\bf S}(x,y)$ (red). (a) A single plane wave propagating along the $x$-axis. (b) Interference of several plane waves with the same frequencies and amplitudes but different directions of the wavevectors.}
\label{Fig_1}
\end{figure*}

In this work, we report on the first controllable generation of these topological structures in gravity water waves. We deal with two basic configurations. First, a suitable interference of three plane waves produces a lattice of: (i) the first-order vortices in the surface-elevation (vertical displacement) field, (ii) skyrmions in the instantaneous 3D surface-displacement field, and (iii) C-points of pure-circular `polarizations' (3D trajectories of water-surface particles) surrounded by the polarization M\"{o}bius strips. Second, circularly-distributed multiple interfering waves with azimuthal phase increment produce Bessel-type wave vortices with controllable topological charges. Most importantly, we demonstrate efficient manipulation of macroscopic floating particles using structured water waves. In complete analogy with optical and acoustic forces and torques, we observe: (i) the gradient force trapping particles in the high-wave-intensity areas, (ii) the `radiation-pressure' force pushing the particle along the local phase gradient, i.e., the wave momentum density, and (iii) the torque on the particle produced by the effective spin density in the water-wave field.

Our results provide an efficient toolbox for smart water-wave control and manipulation of floating objects, and offer a new accessible platform for studies of topological and dynamical properties of structured waves.  

\section{Basic notions}

To begin with, we introduce the main concepts used in this work and shown in Fig.~\ref{Fig_1}. We deal with linear deep-water gravity waves, but all the main results remain valid for gravity-capillary finite-depth waves \cite{Smirnova2024PRL}. In the linear approximation, the water-surface particles oscillate in space and time, which can be characterized by the {\it 3D displacement} of each particle with respect to its unperturbed position, ${\mathbfcal R} (x,y,t)$. The $z$-component of this displacement, ${\mathcal Z}(x,y,t)$, describes the directly observable perturbed water surface. 

Consider first a plane gravity wave with frequency $\omega$ and wavevector ${\bf k} = k \bar{\bf x}$ (the overbar denotes the unit vector of the corresponding direction), related as $\omega = \sqrt{gk}$ ($g$ is the gravitational acceleration), Fig.~\ref{Fig_1}(a). In such a wave, the water-surface particles move along circular trajectories traced by ${\mathbfcal R} (x,y,t)$ in the $(z,x)$ plane \cite{Falkovich_book}. These circular trajectories and the corresponding local angular momentum produced by circling particles can be considered as fluid-mechanical analogues of circular {\it polarizations} and {\it spin} angular momentum density in electromagnetic or acoustic waves \cite{Jones1973, Bliokh2022SA, Bliokh2015PR, Shi2019NSR}.  

In a plane wave, the circular `polarizations' (trajectories) and the corresponding spin vectors are similar for all water-surface particles. However, when we interfere several plane waves with the same frequency but different wavevectors, the behaviour of particles in different points of the $(x,y)$ plane varies dramatically, Fig~\ref{Fig_1}(b). The trajectory of each particle is now a generic ellipse in 3D space, with the corresponding spin normal to its plane. Such single-frequency structured wavefield is conveniently described by the complex displacement field ${\bf R}(x,y)$: ${\mathbfcal R} (x,y,t) = {\rm Re}[{\bf R} (x,y) e^{-i\omega t} ]$. The corresponding time-averaged spin angular momentum density is given by ${\bf S} = (\rho \omega/2) {\rm Im} ({\bf R}^* \times {\bf R})$, where $\rho$ is the water density \cite{Bliokh2022SA, Smirnova2024PRL}. Below we study topological structures in gravity waves in terms of the linear surface-displacement field ${\mathbfcal R}$ or ${\bf R}$, and its physically-meaningful quadratic forms, such as the spin density ${\bf S}$. 

\section{Generation of vortices, skyrmions, and M\"{o}bius strips \\ in three-wave interference}

Our experiments were preformed in a square wave tank with a size of $60\,{\rm cm} \times 60\,{\rm cm}$ and a depth of $h=3\,$cm. We worked with the wavelengths $\lambda = 2\pi/k = 2$--$4\,{\rm cm}$ satisfying the deep-water approximation $\tanh (kh) \simeq 1$.

\begin{figure*}[t]
\centering
\includegraphics[width=0.85\linewidth]{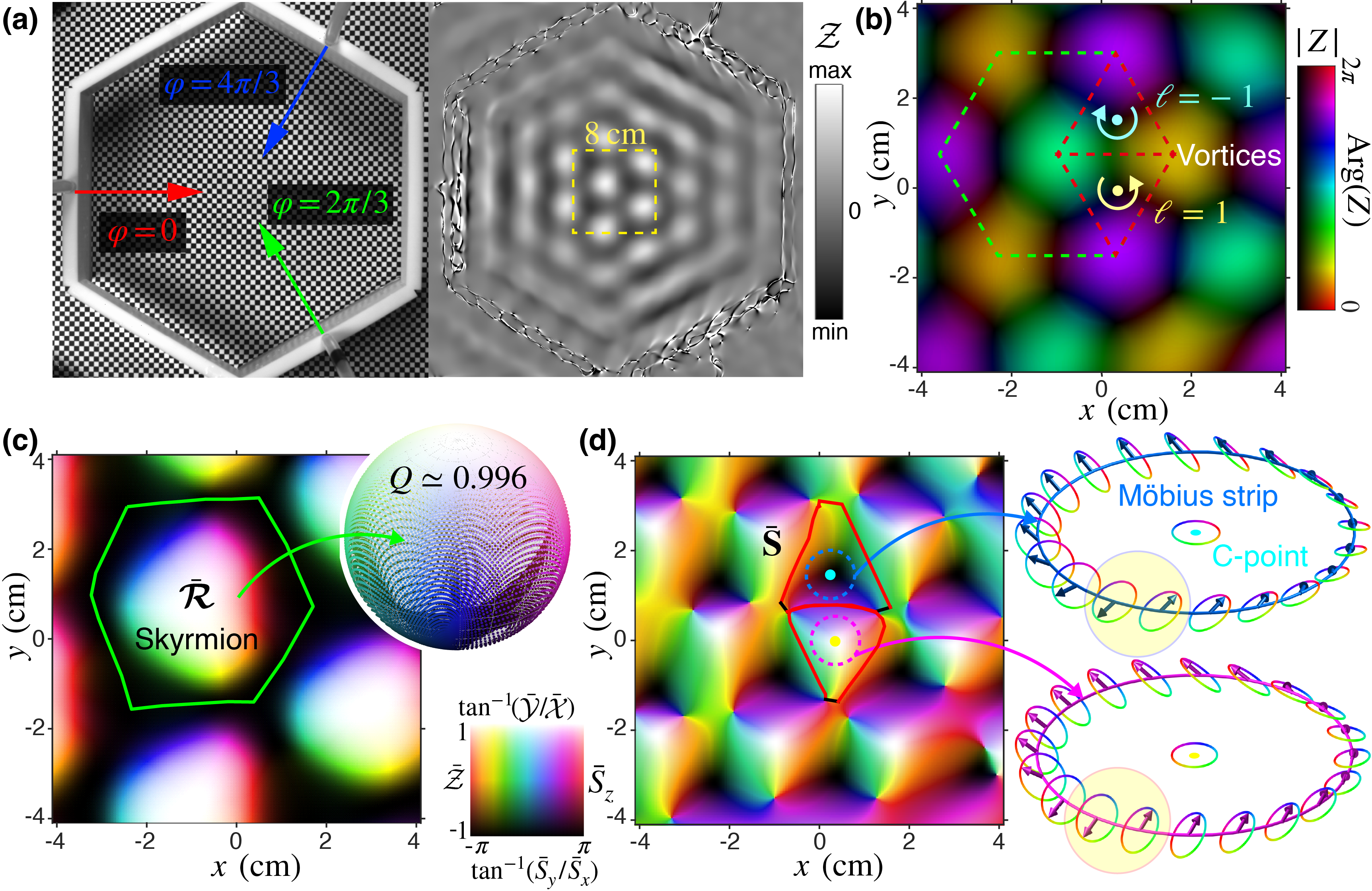}
\caption{{\bf Topological structures in the interference of three water waves.} (a) Experimental setup generating three plane waves with the same amplitude and frequency but different azimuthal angles or propagation $\phi = (0, 2\pi/3, 4\pi/3)$ and the corresponding phases delays $\varphi = \phi$. The measured vertical displacement field ${\mathcal Z}(x,y,t=0)$ is shown on the right. (b) Distributions of the amplitude (brightness) and phase (color) of the corresponding complex field $Z (x,y)$ in the rectangular area indicated by yellow square in (a). Examples of phase singularities (wave vortices) with topological charges $\ell=\pm 1$ are shown by dots and circular arrows indicating the $2\pi \ell$ phase increments around these. (c) Distribution of the reconstructed 3D displacement field ${\mathbfcal R} (x,y,t=0)$ (encoded by the brightness and colors) exhibits skyrmions: the field directions in the near-hexagonal cell are mapped onto the unit sphere with the topological number $Q = 1$. The measured skyrmion boundary is noticeably perturbed as compared to the ideal hexagon shown in (b) but the topological charge $Q$ calculated from the discrete experimental data is very close to the integer theoretical value. (d) The corresponding distribution of the spin density ${\bf S}(x,y)$ exhibits a lattice of near-triangular `failed' merons (half-skyrmions): the meron boundaries determined by $S_z (x,y)=0$ (red curves) are not closed, unlike the ideal triangles in (b). Nonetheless, these quasi-merons contain stable topological structures in the distributions of elliptical polarizations (trajectories) traced by the displacement field ${\mathbfcal R} (x,y,t)$. These are C-points of purely circular polarizations surrounded by the polarization M\"{o}bius strips: orientations of the major semiaxes of the polarization ellipses (blue and magenta vectors) flip (highlighted in yellow) when continuously encircling the C-point.}
\label{Fig_2}
\end{figure*}

In the first experiment, we interfered three plane waves with equal frequencies and amplitudes, but different directions ${\bf k}_i = \bar{\bf x} \cos{\phi_i} + \bar{\bf y} \sin{\phi_i}$, $\phi_i = 2\pi (i-1)/3$, $i=1,2,3$ and phases $\varphi_i = \phi_i$, as shown in Fig.~\ref{Fig_2}(a).
This interference field was produced within a 3D-printed hexagonal structure with sides of $16\,$cm width, where three sides 1,3,5 acted as independent plane-wave sources, while the opposite sides 2,4,6 were open boundaries (raised above the water surface). The whole structure was surrounded by sponge absorbers to avoid wave reflections. The source sides were connected via tubing to speakers controlled by a multi-channel sound card (Presonus ``Quantum 4848''), externally interfaced with a computer. 
We used a sinusoidal signal with a frequency of $\omega/2\pi = 6.79\,$Hz corresponding to the wavelength $\lambda = 4\,$cm.  

To quantify the generated wavefield, we measured the water-surface elevation ${\mathcal Z}(x,y,t)$ using the Fast Checkerboard Demodulation (FCD) technique \cite{Wildeman2018}. 
Namely, a checkerboard pattern (with $0.5\times 0.5\,{\rm cm}^2$ black and white squares) was placed at the bottom of the transparent wave tank. A high-resolution video camera (Andor Zyla 5.5, $2160\times 2560$~pixels, 100~fps), paired with Canon 24-70~mm F2.8L lens, was placed about 1.2~m above the wave tank and recorded the checkerboard pattern distortions caused by the wave. When compared to the undistorted reference pattern, a demodulation algorithm allows recovering the water-surface profile ${\mathcal Z}(x,y,t)$ (see Supplementary Materials). 

Figure~\ref{Fig_2}(a) shows the measured three-wave interference field ${\mathcal Z}(x,y,t=0)$ inside the hexagonal cavity (see Supplementary Video 1 for its temporal evolution). The corresponding complex field $Z(x,y)$ was then obtained using the Hilbert transform, Fig.~\ref{Fig_2}(b). Notably, this field exhibits a hexagonal lattice of {\it phase singularities (wave vortices)} with alternating topological charges $\ell=\pm 1$ \cite{Nye1974, Soskin2001PO, Dennis2009PO, Shen2019LSA, Guo2022JAP, Bliokh2017PR, Smirnova2024PRL}. These are points where the wave amplitude vanishes, $|Z|=0$, whereas the phase ${\rm Arg}(Z)$ increases by $2\pi \ell$ when encircling the point in the counterclockwise direction in the $(x,y)$ plane. Such wave vortices are the simplest topological entities in a scalar structured wavefield; these are analogous to quantized vortices in quantum fluids rather than to classical hydrodynamical vortices \cite{Smirnova2024PRL}. For water waves, square lattices of alternating first-order vortices have been generated by interfering two orthogonal $\pi/2$ phase-shifted standing waves \cite{Filatov2016PRL, Francois2017NC, Bliokh2022SA}.

Our work requires the full-vector wavefield ${\mathbfcal R} (x,y,t)$. Although its direct measurement are challenging, the horizontal in-plane components can be unambiguously reconstructed from the measured vertical component ${\mathcal Z}$ using the gravity-wave equations: $(X,Y) = k^{-1} {\bm \nabla}_2 Z$, where ${\bm \nabla}_2 = (\partial_x, \partial_y)$ is the in-plane gradient. This method is similar to the reconstruction of acoustic vector velocity field from the gradient of the measured scalar pressure field \cite{Shi2019NSR,Muelas2022PRL}.

Figure~\ref{Fig_2}(c) shows the color-brightness-coded distribution of the instantaneous 3D displacement field ${\mathbfcal R} (x,y,t=0)$ reconstructed from the three-wave interference measurements. Remarkably, it exhibits a hexagonal lattice of {\it skyrmions}: areas where the directions of unit vectors $\bar{\mathbfcal R}$ can be mapped onto the unit sphere, with opposite directions $\bar{\mathbfcal R} = \bar{\bf z}$ in the center and $\bar{\mathbfcal R} = - \bar{\bf z}$ in the vertexes. The topological number of this mapping is $Q = (1/4\pi) \iint \bar{\mathbfcal R} \cdot (\partial_x \bar{\mathbfcal R} \times \partial_y \bar{\mathbfcal R})\, dx dy = 1$ at $t=0$, and its sign flips after every half-period of the wave oscillations. Akin to vortices, skyrmions are topologically protected structures. In our experiment, the hexagonal skyrmion boundary is noticeably perturbed (we determine it as a local-minimum curve for $\bar{\mathcal Z} (x,y,0)$) but the calculated topological number $Q \simeq 1$ with an accuracy of $10^{-3}$.
To the best of our knowledge, this is the first observation of water-wave skyrmions, similar to the ones observed recently in optical \cite{Tsesses2018S, Deng2022NC, Shen2024NP} and acoustic \cite{Ge2021PRL, Muelas2022PRL, Cao2023SA} waves.

\begin{figure*}[t]
\centering
\includegraphics[width=0.8\linewidth]{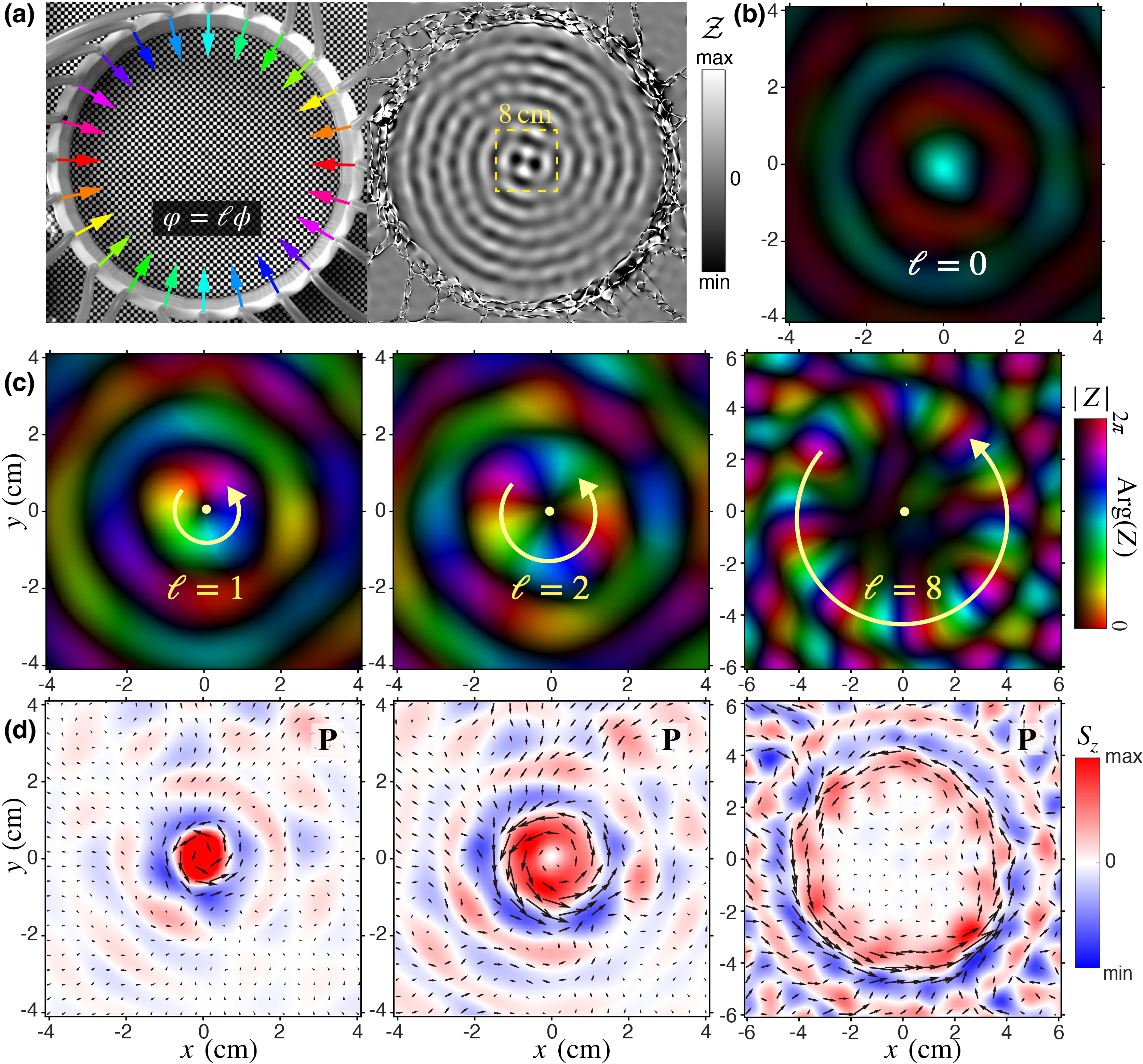}
\caption{{\bf Generation of the Bessel-type water-wave vortices with different topological charges.} (a) Experimental setup with $N=24$ sources uniformly distributed around a circle with azimuthal phase difference $\varphi = \ell \phi$ corresponding to the integer topological charge $\ell$ (here $\ell=2$ is shown). The measured vertical displacement field ${\mathcal Z}(x,y,t=0)$ is shown on the right. (b,c) Distributions of the amplitude (brightness) and phase (color) of the reconstructed complex field $Z(x,y)$ for $\ell=0,1,2,8$. (d) Distributions of the time-averaged vertical spin density $S_z (x,y)$ (color density plots) and in-plane wave momentum density ${\bf P}(x,y)$ (black vectors) for the Bessel vortices shown in (c).}
\label{Fig_3}
\end{figure*}

Finally, we analyze the local elliptical polarizations (trajectories) traced by the experimentally retrieved displacement field ${\mathbfcal R} (x,y,t)$ and the corresponding spin density ${\bf S}(x,y)$. The distribution of the normalized spin vectors $\bar{\bf S}(x,y)$ is shown in Fig.~\ref{Fig_2}(d). For an ideal three-wave interference, they form a lattice with triangular {\it meron (half-skyrmion)} cells \cite{Dai2020N, Shen2024NP, Smirnova2024PRL}. Each triangular cell (1/6 of the hexagonal cell) is centered around the phase singularity of the field $Z(x,y)$, where the spin is directed vertically: $\bar{\bf S} = \ell\, \bar{\bf z}$, whereas the boundaries correspond to purely horizontal spin: ${S}_z =0$. Such merons are mapped onto the upper or lower semisphere of spin directions with the corresponding topological numbers $Q_S = \ell/2$. However, these spin merons are {\it not} topologically stable, because the vertexes of the triangles correspond to higher-order singular points with ${\bf S}={\bf 0}$. In a real perturbed system, such as our experiment, these singular points split, the perturbed ${S}_z =0$ lines do not form closed boundaries and the spin merons fail, see Fig.~\ref{Fig_2}(d) and Supplementary Materials.    

Instead of the spin merons, the elliptical polarizations ${\mathbfcal R} (x,y,t)$ exhibit another kind of topologically stable structures. Namely, centers of the broken spin-meron triangles contain {\it polarization singularities}: {\it C-points} with purely circular polarizations \cite{Nye1987, Dennis2009PO, Muelas2022PRL, Bliokh2021POF}. Orientations of the major semiaxes of the 3D polarization ellipses around the C-points form {\it polarization M\"{o}bius strips}: the semiaxis direction flips when continuously encircling the polarization singularity \cite{Freund2010OC, BAD2019}. These are robust topological structures, previously observed in optical \cite{Bauer2015S} and acoustic \cite{Muelas2022PRL} fields, and recently predicted theoretically \cite{Bliokh2021POF} for water waves.
Figure~\ref{Fig_2}(d) shows the polarizations M\"{o}bius strips around two C-points (approximately corresponding to phase singularities of $Z(x,y)$) retrieved from our experimental measurements of the water-wave interference field. 


\section{Generation of Bessel-type vortices with different topological charges}

In the second experiment, we demonstrated the generation of water-wave vortex modes with different topological charges. For this goal, we interfere $N=24$ waves with azimuthal directions uniformly distributed around a circle, $\phi_i = 2 \pi (i-1) / N$, $i=1,...,N$, and mutual phase delays corresponding to the vortex of a topological charge $\ell$: $\varphi_i = \ell \phi_i$, $\ell=0,\pm 1, \pm 2, ...$., Fig.~\ref{Fig_3}. In the limit of $N \gg 1$ this interference produces a circularly-symmetric Bessel-type vortex field \cite{Smirnova2024PRL}. We used a 3D-printed 24-gon structure in the same wave tank, which approximated a circle with a radius of $18\,$cm, Fig.~\ref{Fig_3}(a). The 24 sides were connected via tubing to speakers controlled by a multi-channel sound card, providing coherent sources with computer-controlled amplitudes and phases. 
We used sinusoidal signal with a frequency of $\omega/2\pi = 8.66\,$Hz corresponding to the wavelength $\lambda = 2\,$cm. The diameter of the near-circular structure accommodated an integer number of wavelengths, so that each pair of oppositely-directed waves formed a stable standing wave within the structure. 

Akin to the three-wave interference experiment, we measured the vertical displacement field ${\mathcal Z} (x,y,t)$ via the FCD technique. An example of the measured field at $t=0$ for $\ell =2$ is shown in Fig.~\ref{Fig_3}(a). The corresponding complex fields $Z(x,y)$ for $\ell=0, 1, 2, 8$ are shown in Figs.~\ref{Fig_3}(b,c) (see also Supplementary Videos 2--5 for temporal evolutions of the  measured fields ${\mathcal Z} (x,y,t)$). These fields correspond to the Bessel-type vortices $Z\propto J_{|\ell|}(kr)\exp (i\ell\phi)$ ($J_{|\ell|}$ is the Bessel function of the first kind) with $2\pi\ell$ phase increment around the center \cite{Smirnova2024PRL}. (Notably, the 3D displacement field ${\mathbfcal R}(x,y,t=0)$ of the $\ell=0$ non-vortex mode forms a skyrmion in the center, see Supplementary Materials.) For $\ell \neq 0$, these are quasi-standing waves which do not propagate in the radial direction, but do propagate in the azimuthal direction. The wave amplitude is maximum near the first Bessel-maximum ring of radius $r_{\rm max} \simeq  \sqrt{|\ell|}\lambda/2$. To the best of our knowledge, this is the first controllable generation of higher-order wave vortices in water waves. 

Such wave vortices exhibit remarkable dynamical properties carrying both spin and OAM, so that the $z$-component of the total (spin + orbital) angular momentum is quantized according to the topological number $\ell$ \cite{Smirnova2024PRL}. To quantify these angular-momentum properties, we reconstruct the complex 3D displacement field ${\bf R}(x,y)$, as in the three-wave experiment. Then we calculate the corresponding spin density ${\bf S}(x,y)$, as well as the canonical wave {\it momentum density} ${\bf P} (x,y) = (\rho \omega /2)\, {\rm Im} [{\bf R}^* \cdot ({\bm \nabla}_2) {\bf R}]$. This wave momentum density is directly related by the velocity ${\bf U}_S$ of the {\it Stokes drift} of water-surface particles, ${\bf P} =\rho\, {\bf U}_S$, which appears as a time-averaged nonlinear quadratic correction to their linear oscillatory motion \cite{Bremer2017, Falkovich_book, Francois2017NC, Bliokh2022SA}. The $z$-component of the OAM density is described by the azimuthal component of the momentum density: $L_z = ({\bf r} \times {\bf P})_z = r P_\phi$ \cite{Smirnova2024PRL}. 

Figure~\ref{Fig_3}(d) displays the distributions of the vertical spin density $S_z$ and the wave momentum density ${\bf P}$ calculated from the experimentally measured Bessel-vortex fields with $\ell=1,2,8$. One can clearly see strong azimuthal momentum $P_\phi$ around the first Bessel-maximum ring, as well as rings of positive and negative spin $S_z$ around it. 
(Note that similar distributions of the wave momentum density were measured via the Stokes-like drift of fluid particles induced by acoustic Bessel beams \cite{Hong2015PRL}.) 
Flipping the vortex sign ${\rm sgn}(\ell)$ results in the sign flipping for both the azimuthal momentum $P_\phi$ and spin $S_z$. For the first-order vortices, $|\ell|=1$, the spin density $S_z$ reaches its maximum/minimum in the vortex center $r=0$, similar to the vertical-spin extrema near the first-order vortices and C-points in the three-wave interference, Fig.~\ref{Fig_2}. Below we reveal dynamical manifestations of these momentum and spin properties in the interaction of water-wave vortices with floating particles.

\section{Trapping and manipulation of floating particles}

In this series of experiments we used subwavelength spherical polyethylene particles with density $\rho_p =  0.89\!-\!0.96\,$g/cm$^3$ and radii $a=$~2.4, 3.1, 4.75, 6.35~mm, floating on the water surface. We also used a standard ping-pong ball with $a=20\,$mm. 
Such floating particles undergo fast oscillatory motions with the wave frequency $\omega$ together with water-surface elements. In addition, they experience a slow {\it time-averaged} action of the wavefield, {\it quadratic} in the field amplitude. This action is similar to optical and acoustic radiation forces and torques, which underpin optical and acoustic trapping and manipulations \cite{Grier2003N, Gao2017LSA, Ozcelik2018NM, Dholakia2020NRP}.  

We first introduce a theoretical toy model of water-wave-induced force and torque on a floating particle, employing the analogy with optical and acoustic forces and torques in structured monochromatic waves \cite{Toftul2019PRL, Bliokh2015PR, Bliokh2014NC}. This model is based on the main time-averaged dynamical properties of gravity water waves: the wave momentum density ${\bf P}$, the spin density ${\bf S}$, and the kinetic energy density (intensity) $W=\rho \omega^2 |{\bf R}|^2/4$. 
Assuming the lowest-order dipole-like coupling between the water-wave field and the particle, quantified by the complex scalar `polarizability' coefficient $\alpha$, the wave-induced force and torque on the particle can be written as (see Supplementary Materials):
\begin{equation}
{\bf F} = {\rm Re}(\alpha) {\bm \nabla}_2 W + \omega\, {\rm Im} (\alpha)\, {\bf P} \equiv {\bf F}_{\rm grad} + {\bf F}_{\rm press}\,,\quad 
T_{\rm spin} = \omega\, {\rm Im} (\alpha)\, S_z\,.
\label{force}    
\end{equation}
Here we describe the action affecting the 2D motion of the particle in the $(x,y)$ plane, i.e., the horizontal in-plane force and vertical $z$-directed torque. The first term in the force (\ref{force}) is the {\it gradient force} responsible for the trapping of particles in high-intensity or low-intensity zones [depending on the sign of ${\rm Re}(\alpha)$], while the second term is the {\it wave-pressure force} directed along the local wave momentum. The wave-pressure force and spin-induced torque are directly related to the transfer of the local momentum and spin angular momentum from wave to particle \cite{Bliokh2015PR, Bliokh2014NC, Toftul2019PRL}.

\begin{figure*}[t]
\centering
\includegraphics[width=0.85\linewidth]{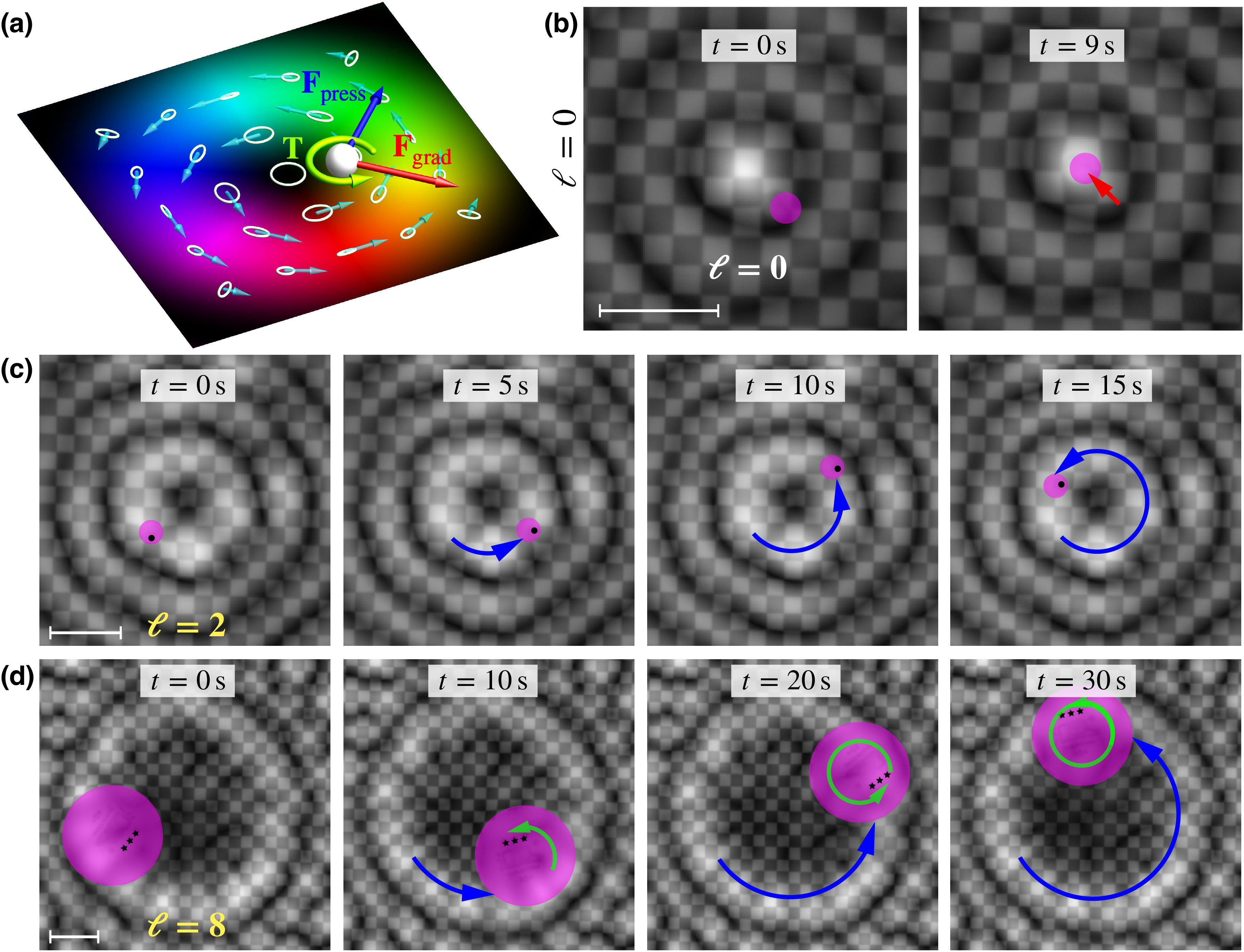}
\caption{{\bf Dynamics of floating particles in the Bessel-vortex water waves.} (a) Schematics of the radial gradient and azimuthal wave-pressure forces, together with the spin-induced torque, described by Eqs.~(\ref{force}). (b) Trapping of the $a=2.4\,$mm particle in the central intensity maximum of the $\ell=0$ mode provided by the radial gradient force. (c) Radial trapping of the $a=3.1\,$mm particle in the first Bessel maximum of the $\ell= 2$ vortex, accompanied by its orbital motion due to the azimuthal wave-pressure force (cf. the wave-momentum distribution in Fig.~\ref{Fig_3}(d)). (d) Similar trapping and orbital motion of a ping-pong ball, $a=20\,$mm, in the $\ell= 8$ vortex. The ball also experiences spinning rotation (green arrows and the three-star markers) induced by the torques due to the spin density $S_z$ and due to the radial gradient of the azimuthal wave-pressure force. In panels (b--d), for better visibility, the actual video frames of floating particles with the checkerboard background are overlapped with the experimentally measured wave amplitude $|Z|(x,y)$ (grayscale), whereas the particle position and orientation are highlighted by the magenta and black markers. The scale bars correspond to 2~cm.}
\label{Fig_4}
\end{figure*}

In optics and acoustics, in the small-particle limit $ka\ll 1$, the polarizability typically scales proportionally to the particle's volume: $\alpha \propto a^3$, and the imaginary part ${\rm Im} (\alpha)$ characterizes absorption of waves by the particle. Therefore, the wave-pressure force and spin-induced torque in Eq.~(\ref{force}) can be associated with the transfer of the momentum and spin angular momentum from the water-wave field to the particle.  

\begin{figure*}[t]
\centering
\includegraphics[width=0.85\linewidth]{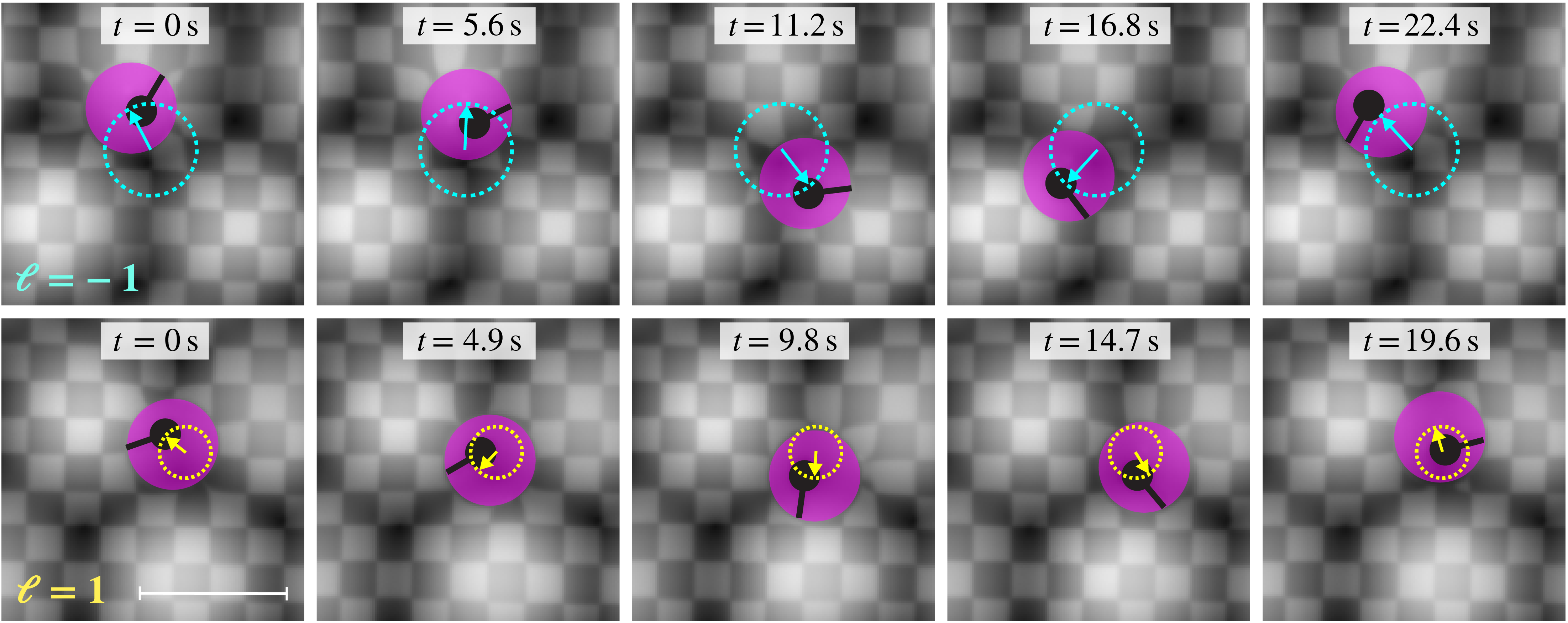}
\caption{{\bf Dynamics of floating particles around the first-order vortices in the three-wave interference lattice.} Particles of radius $a=6.35\,$mm are placed near the $\ell=\pm 1$ vortices shown in Figs.~\ref{Fig_2}(b,d). The particles are trapped within the triangular quasi-spin-meron zones, experiencing orbital and spinning rotations. These rotations are non-uniform because of the circularly-asymmetric character of the vortices and are shown via approximate circular trajectories of the particle's center (dashed circles) and black markers indicating the particle's orientation.}
\label{Fig_5}
\end{figure*}

Figure~\ref{Fig_4}(a) schematically shows the action of the gradient force, the wave-pressure force, and the spin-induced torque, Eqs.~(\ref{force}), on a particle in a water-wave vortex field. The gradient and wave-pressure forces are directed radially and azimuthally, according to the intensity and phase gradients in this field, while the torque is induced by local elliptical polarizations (trajectories) of water-surface particles in the $(x,y)$ plane.

Figures~\ref{Fig_4}(b--d) (see also Supplementary Videos 6--9) display the observed dynamics of floating particles of different sizes in the Bessel-vortex wavefields with different topological charges $\ell$, Fig.~\ref{Fig_3}. First, Fig.~\ref{Fig_4}(b) shows that a particle with $a=2.4\,$mm is attracted to the center of the $\ell=0$ Bessel wave and is {\it trapped} there (Supplementary Video 6). This is a clear evidence of the radial gradient force corresponding to ${\rm Re}(\alpha) >0$. (This is somewhat counterintuitive because in acoustics ${\rm Re}(\alpha) < 0$ for $\rho_p < \rho$.) Second, for the vortex with $\ell=2$, the particle with $a=3.1\,$mm is trapped in the first Bessel-maximum ring, and we observe its {\it orbital motion} due to the azimuthal wave-pressure force, Fig.~\ref{Fig_4}(c) (Supplementary Video 7). This is a direct mechanical confirmation of the OAM $L_z$ carried by water-wave vortices and transferred to the particle. (The particle reaches a constant angular velocity when the azimuthal pressure force is balanced by the friction proportional to the velocity.) Furthermore, we observe similar radial trapping and orbital motion for a ping-pong ball ($a=20\,$mm) in a Bessel vortex with $\ell=8$, Fig.~\ref{Fig_4}(d) (Supplementary Video 9). Notably, in addition to the orbital motion, the ping-pong ball also experiences a {\it spinning rotation} with respect to its center. The angular velocities of the spinning and orbital rotations approximately equal $\Omega_{\rm spin} \simeq 2\, \Omega_{\rm orb} \simeq \pi/10\,$rad/s ($\ll \omega$). The spinning rotation is caused by the wave-induced torque with respect to the ball's center. This torque can have two origins: (i) the spin-induced torque in Eq.~(1) [note the positive spin density $S_z>0$ at the inner part of the Bessel ring, where the ball is trapped, Fig.~\ref{Fig_3}(d)] and (ii) the `gradient torque' $T_{\rm grad}$ originated from the radial gradient of the azimuthal wave-pressure force, which pushes the outer side of the ball stronger. Both of these torques act in the same direction and contribute to the spinning rotation (see Supplementary Materials). 

Remarkably, this observed dynamics is entirely analogues to the radial trapping, orbital, and spinning motion of small particles observed in optical vortex fields \cite{ONeil2002PRL, Garces2003PRL}. The only difference is that in optics the spinning motion can be controlled by the global circular polarization of the wave, independently of the vortex charge $\ell$, while in water waves, the signs of the orbital and spinning rotations are locked with each other. Note also that the radial gradient force are not always sufficient for stable trapping: particles of certain sizes can escape the vortex ring when the centrifugal force prevails (see the case of $a=4.75\,$mm particle in the $\ell=2$ vortex in Supplementary Video 8). This effect can be employed for accurate measurements of the trapping force. 

We finally show that the above water-wave manipulations of floating particles are quite robust and do not require circularly-symmetric Bessel vortices. For this purpose, we place particles of radius $a=6.35\,$mm near the centers of the $\ell=\pm 1$ vortices in the three-wave interference lattice, Fig.~\ref{Fig_2}. Figure~\ref{Fig_5} (see also Supplementary Videos 10 and 11) shows that the particles are trapped in around such vortices, within the quasi-spin-meron traingular zones. Furthermore, the particles undergo both orbital and spinning rotations controlled by ${\rm sgn}(\ell)$. In this case, the angular velocity of the orbital motion is larger: $\Omega_{\rm obr} \simeq 2\,\Omega_{\rm spin} \simeq \pi/10\,$rad/s. The spinning motion can be induced by the two kinds of torques mentioned above, particularly by the maximum/minimum spin density $S_z$ in the centers (C-points) of the quasi-spin-meron zones, Fig.~\ref{Fig_3}(d) (see Supplementary Materials).    

\section{Conclusions}

We have demonstrated the controllable generation of topological water-wave structures, including vortices of different topological charges, skyrmions, and polarization M\"{o}bius strips. Our experiments evidence the robustness of these structures, in contrast, e.g., to spin merons `broken' in a perturbed three-wave interference. Most importantly, we have revealed remarkable dynamical properties of these topologically structured water waves and demonstrated their capability of manipulating floating particles of different sizes, including stable trapping, as well as orbital and spinning rotations. 

Thus, our results offer a new platform for wave-induced mechanics, which can extrapolate the well-developed optical and acoustic manipulations to fluid mechanics. In particular, capillary water-surface waves can be employed for microfluidic manipulation of biomedical objects similar to the presently used acoustic waves \cite{Ozcelik2018NM, Dholakia2020NRP, Ding2013LC}. While optical manipulations work with wavelengths of the order of micron, acoustic manipulations occupy the wavelength range from the 10s of microns to millimeters, water waves can efficiently work in the next range from millimeters to centimeters and further to colossal ocean waves. 

This work is only the first step in this direction. Further development can include a detailed consideration of the interaction between structured water waves and floating particles, sorting of particles with different properties using water waves, the interplay between topology and  nonlinearity inherent in water waves, consideration of other topological structures, multi-frequency, and complex spatio-temporal waves, and so forth. 


\vspace{0.5cm}

\begin{acknowledgements}

{\bf Acknowledgements:}
This work was partially supported by 
National Key Research and Development Program of China
(2022YFA1404800, 2021YFA1400603); National Natural Science Foundation of China (No. 12234007, No. 12221004, and No. 12104132); Major Program of National Natural Science Foundation of China (No. 91963212); Science and Technology Commission of Shanghai Municipality (22142200400, 21DZ1101500, 2019SHZDZX01); China Postdoctoral Science Foundation (2022M720810, 2022TQ0078, and 2023M741024), Nanyang Technological University Start Up Grant, Singapore Ministry of Education (MOE) AcRF Tier 1 grant (RG157/23), MoE AcRF Tier 1 Thematic grant (RT11/23), Imperial-Nanyang Technological University Collaboration Fund (INCF-2024-007),
 the International Research Agendas Programme (IRAP) of the Foundation for Polish Science co-financed by the European Union under the European Regional Development Fund and Teaming Horizon 2020 program of the European Commission [ENSEMBLE3 Project No. MAB/2020/14], and the project of the Ministry of Science and Higher Education (Poland) “Support for the activities of Centers of Excellence established in Poland under the Horizon 2020 program” [contract MEiN/2023/DIR/3797].

{\bf Author contributions:}
Y. S., L. S., K. Y. B., and J. Z. conceived the basic idea of the work and supervised its development. 
Z. C., and L. S. designed the experiments.
Z. C. performed the experiments. 
B. W. and Z. C. analyzed the experimental data. 
K. Y. B. prepared the first version of the manuscript with input from all the authors. 
All authors took part in discussions, interpretations of the results, and revisions of the manuscript.

{\bf Competing interests:} 
The authors declare no competing interests.

{\bf Data and materials availability:} 
All data needed to evaluate the conclusions in the paper are present in the paper and the Supplementary Materials.
   
\end{acknowledgements}

%

\end{document}